# Does an awareness of differing types of spreadsheet errors aid end-users in identifying spreadsheets errors?

Michael Purser, David Chadwick
University of Greenwich, Old Royal Naval College, Greenwich, London SE10 9LS
michael-purser@beeb.net

**ABSTRACT**

The research presented in this paper establishes a valid, and simplified, revision of previous spreadsheet error classifications. This investigation is concerned with the results of a web survey and two web-based gender and domain-knowledge free spreadsheet error identification exercises. The participants of the survey and exercises were a test group of professionals (all of whom regularly use spreadsheets) and a control group of students from the University of Greenwich (UK). The findings show that over 85% of users are also the spreadsheet's developer, supporting the revised spreadsheet error classification. The findings also show that spreadsheet error identification ability is directly affected both by spreadsheet experience and by error-type awareness. In particular, that spreadsheet error-type awareness significantly improves the user's ability to identify, the more surreptitious, qualitative error.

## 1      INTRODUCTION

This paper is an investigation into the level and effect of spreadsheet error-type awareness within the professional spreadsheet user community. The various sections contained within this paper are briefly outlined in the following paragraphs.

Section 2 outlines the reported issues of spreadsheet integrity and what has been done to alleviate these problems. Previous and current research into the classification of spreadsheet errors is considered and a spreadsheet error-type classification suggested. Section 3 details the investigation of spreadsheet error-type awareness; the research questions and hypotheses; the procedures and participants; and the methods employed. Section 4 discusses the quantitative and qualitative analysis of the results of the investigation. Section 5 draws conclusions based upon the results of the investigation. Section 6 considers what further research is necessary into spreadsheet model integrity.

## 2      SPREADSHEET PROBLEMS

For the world's financial organisations the ultimate killer desktop application is the ubiquitous spreadsheet. Commonly known – though unpublished and intentionally unconfirmed by Microsoft – Excel has an install base of 90% on end-user desktops. From its humble beginnings as VisiCalc to its dominance as Excel, the spreadsheet has served as a repository for financial models and consequently as a decision-making support tool for over two and a half decades.

A recent article by Grenville J. Croll (2005), from the 2005 European Spreadsheet Risk Interest Group (EuSpRIG) conference, quoted the Corporation of London's web site as stating that the City of London's international financial markets contributed £13bn to the UK's current account.

Grenville Croll also indicated that this was supported by approximately 317,000 "*City-Type*" financial jobs and,

> "...estimate that up to 30% of these jobs would be working with Important, Key or Critical spreadsheets..."

By this calculation there are over 95,000 City of London workers, using important, key or critical spreadsheets, to generate some of the £13bn revenue. This provides ample potential of an aggregation of seemingly insignificant errors into large financial miscalculations.

Spreadsheet errors have continually made the news sections of many of the leading financial journals. There have been regular reports with alarming headlines such as, *BUSINESS MODELS VITAL BUT '95% CONTAIN MAJOR ERRORS'* by the Magazine for CMAs (2005). These articles report how independent auditors, such as KPMG, were finding that

> "...92% of spreadsheet-based models... had significant errors [and] 75% had significant accounting errors...."

In 2005, the activities of the European Spreadsheet Risk Interest Group (EuSpRIG) came to the notice of the Daily Telegraph reporter Robert Miller (2005) in its Money supplement. Robert Miller quotes EuSpRIG's membership secretary Grenville Croll's statement, that

> "...the probability for material error is way over 90%..."

The simple solution would be the replacement of spreadsheets with centralised and hence more easily controlled applications. It isn't that simple though, as another speaker at the EuSpRIG conference, Dr. Buckner pointed out,

> "Spreadsheets are integral to the function and operation of the global financial system. It's unrealistic to expect them to be replaced in the short term."

Auditors and academics have published papers, articles, and guidance on two inter-related areas; spreadsheet error rates, and how to generate error-free, trustworthy and reliable financial spreadsheets.

Resolving the first of these problems, spreadsheet errors, will likely provide the greatest return on investment.

**2.1   Responses to the Problems**

In practice responses to the problem have generally been remedial in their nature. In 2001/2002 BBC news (2002) reported the accounting scandals of Enron, WorldCom and AIB. The BBC reported how these incidents caused the fast track introduction of the Sarbanes Oxley Act (SOX) with the intention (not limited to, but including spreadsheets) that:

> "...the audit process was paramount, particularly in establishing good governance of financial information."

However, SOX has not eradicated occurrence of spreadsheets errors by adherence to an audit process. And, the problem of overcoming spreadsheet errors has continued to gain prominence in the professional journals.

Auditors PwC in an IMA (2004) report focused on a development control and audit approach. It is likely that this approach succeeds where auditing is part of the user's skill set and it is not considered an overhead but integral to any operation. This approach relies heavily upon the auditor's know-how to assist them in identifying errors.

Some authors such as Bellinger (2005), in an article in IT Training suggests that;

> "…the problem is down to a lack of spreadsheet skills."

As spreadsheets are simple to use, very little formal training is given. Yet, spreadsheets can be very complex and it takes many skills to develop a true mastery of modelling with them. Training may be the answer. In the same article, Bellinger hints at another key realisation. The developer and end-user may often be one-in-the-same,

> "as the IT department just shrugs and says that spreadsheets fall outside their jurisdiction."

This may be one of the most important key factors in reducing the occurrence of errors in spreadsheets. The realisation that the end-user and the developer are one-and-the-same solves an issue in the development life cycle approaches taken by spreadsheets academics. That is, the question of who creates the error. Considering a single developer-user, means the 'who' can be removed.

## 2.2 Identifying Errors

Without doubt the most critical success factor in reducing the occurrence of errors is the ability of the application, developer and user to identify an error. Knowing what to look for is the initial step in the process. Thus, it is essential a classification of error-types be developed.

All errors exhibit characteristics. Therefore, each error must exhibit a set of one or more characteristics, and the observation of characteristics common to one or more errors must provide a mechanism for classifying errors into types. Classification of these characteristics may provide classification rules that could be expressed in a taxonomic form. The rules for the taxonomic classification of error-types may be described thus:

Each error-type must exhibit a single set of distinct characteristics, and this distinct characteristics set must not be:

(i) a subset of another error-type's characteristics set, or

(ii) a duplication of another error-type characteristics set,

Otherwise, the error-type should be considered to be:

(i) a child error-type (error-sub-type) of a parent error-type (error-super-type), or,

(ii) the same error-type.

The simplest way of expressing rules in logical and legitimate operations is through the creation of bifurcation tests. Following Rajalingham's (2005) bifurcation test approach and the prototype characteristic rules as a basis, the general rule below is developed and applied recursively to determine the error-type of the error:

> **IF** *error characteristics = error-type characteristics subset*
>
> **THEN** *error is of an error-sub-type of the error-type*
>
> **ELSE IF** *error characteristics = error-type characteristics*
>
> **THEN** *error is of this error-type*
>
> **ELSE** *error is a sibling error-type.*

The characteristics of error-types are beyond the scope of this research. However, in order to understand if error-type awareness impacts upon users' ability to identify and rectify errors, it is first necessary to establish a set of logical and legitimate error-types using the above criteria. It is then possible to classify errors through navigation of the classification, using the above bifurcation test.

## 2.3   Error Types

As previously stated errors may be identified by either, the application, or the developer/user.

The application is able to examine characteristics of an error that the developer/user may not so easily be able to observe, e.g. knowing the data type of the error: text, integer, or fixed decimal. This is because the characteristic may not be immediately visible at the application interface.

The reverse is also true: the application may not be able to examine characteristics of an error that the developer/user may easily be able to observe, e.g. knowing that inappropriate mathematical reasoning has been applied in the spreadsheet or that certain data is no longer valid.

For these reasons it is sensible to consider identification of error-types firstly by the application and secondly by the developer/user – the operator.

**Application Identified Error Types**

Microsoft Excel's Help documents eight application identified error-types, as detailed in table 2.1.

These error types indicate that an error exists that Excel is unable to resolve automatically. The error identification is achieved through a combination of syntactical analysis with argument volume and type analysis. Further detailed explanation of this function is beyond the scope of this research, as this research focuses upon user identified errors.

Each of the eight errors has specific occurrences:

| Error Displayed | Reason for Occurrence |
|---|---|
| #DIV/0! | Occurs when a number is divided by zero (0). |
| #VALUE! | Occurs when the wrong type of argument or operand is used. |
| #NUM! | Occurs with invalid numeric values in a formula or function. |
| #NULL! | Occurs when an intersection of two areas that do not intersect is specified, or the intersection operator is a space between references. |
| #N/A | Occurs when a value is not available to a function or formula |
| #NAME? | Occurs when Microsoft Excel doesn't recognize text in a formula |
| #REF! | Occurs when a cell reference is not valid |
| ##### | Occurs when a column is not wide enough, or a negative date or time is used |

**Table 2.1: Excel's error-types and brief description of causes.**

**Developer/User Identified Error-Types**

Some academic researchers have established classifications of error-types as either the prime aim or a by-product of their spreadsheet research.

A classification of error-types directly related to spreadsheets was developed in 1993 by Galletta *et al.* (1993) in a study of the error finding performance of 60 subjects, CPA (Certified Professional Accountants) and MBA (Masters in Business Administration) graduates. Galletta's 2 dimensional (domain and device) approach is too simplistic and suffered from the flaw of overlapping classifications of error-types.

At the 29th Annual Hawaii International Conference on System Sciences, Panko & Halverson (1996) published their classification of spreadsheet errors in a paper on a survey into spreadsheet risk. The classification was not explicitly taxonomic. However, the classification groups' errors fall into two main types; quantitative and qualitative. In the published paper Panko & Halverson indicate a dichotomy exists between quantitative and qualitative errors in that;

> "...***quantitative*** *errors are numerical errors that lead to incorrect bottom-line values...*"

And,

> "...***qualitative*** *errors, in turn, are flaws that do not produce immediate quantitative errors. But they do degrade the quality of the spreadsheet*

> *model and may lead to quantitative errors during later 'what if' analyses or updates to the model."*

Though considering classification more fully than any previous classifications, Panko & Halverson's 1996 classification does not appropriately describe error examples. Identification of errors using this classification may result in certain errors displaying characteristics from more than one sibling error-type. Thus, it must be concluded that Panko & Halverson's classification of error-types fails to meet the criteria of easily distinguishing between error-types.

Rajalingham, Chadwick and Knight's (2000) paper "Classification of Spreadsheet Errors" presented at the 2000 EuSpRIG conference produced a classification as a result of a thorough investigation of spreadsheet errors. The classification was developed based on the nature and characteristics of errors and the spreadsheet development life cycle.

One of the goals of the classification was to

> *"...minimise the recurrence of the same category or type in different parts of taxonomy."*

While a benefit of the classification was to be that

> *"...any effort to devise a solution or method of detection* [could be applied] *to address errors within the same category of classification."*

The stated approach was to use a binary tree employing dichotomies to classify the spreadsheet errors. And, for the first time it was possible to see that non-human generated errors (i.e. system generated errors) were considered. However, as Rajalingham, Chadwick and Knight stated

> *"...their occurrence is generally beyond the control of users, although they can, when aware, take corrective action."*

Rajalingham (2005) recently made revisions to the original 2000 classification. In a paper published at the 2005 EuSpRIG conference, Rajalingham's classification focused on a classification of

> *"...only user generated spreadsheet errors"*

as system generated errors were beyond the scope of his research, possibly because, as previously stated, they are generally beyond the scope of the user to rectify them. Figure 2.1 shows the revised classification.

As with the previous classifications though, Rajalingham is considering the error creator characteristics by stating that:

> *"Structural errors are produced by the developer of the spreadsheet model."*

And

> *"Data Input errors are made by end-users who merely manipulate the worksheet."*

In considering these actors it could be argued that a potential data input error is actually caused by a structural error when the developer fails to create a robust structure (formula network) in the spreadsheet.

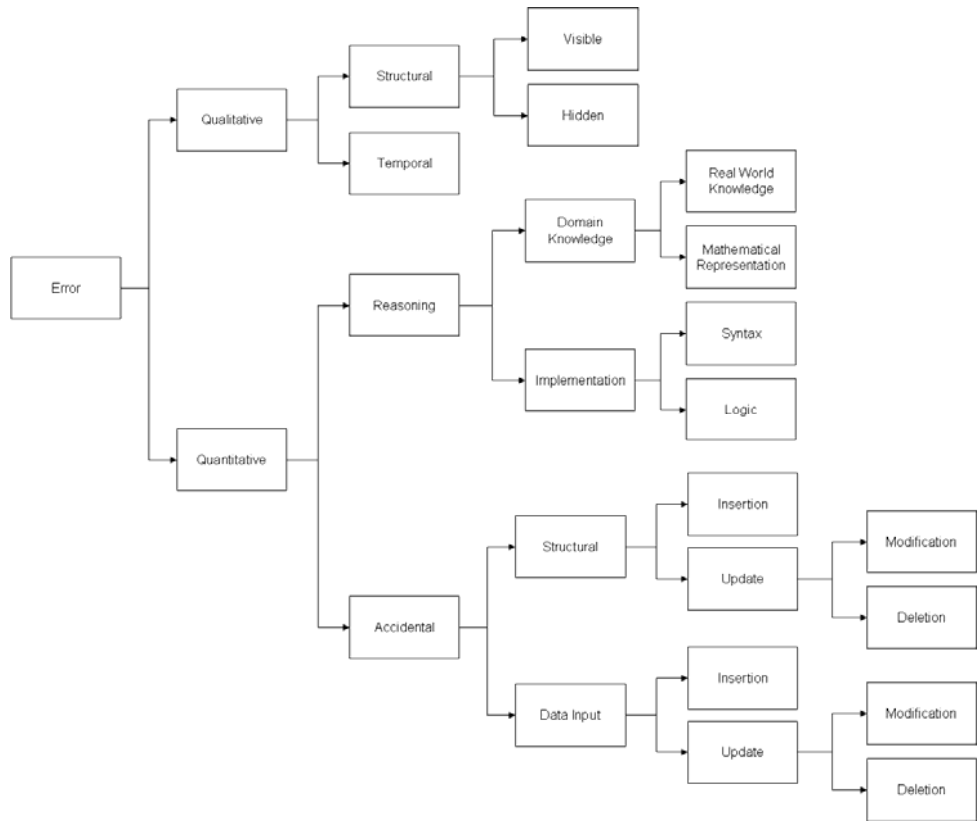

**Figure 2.1: Rajalingham's 2005 revised classification of spreadsheet error-types.**

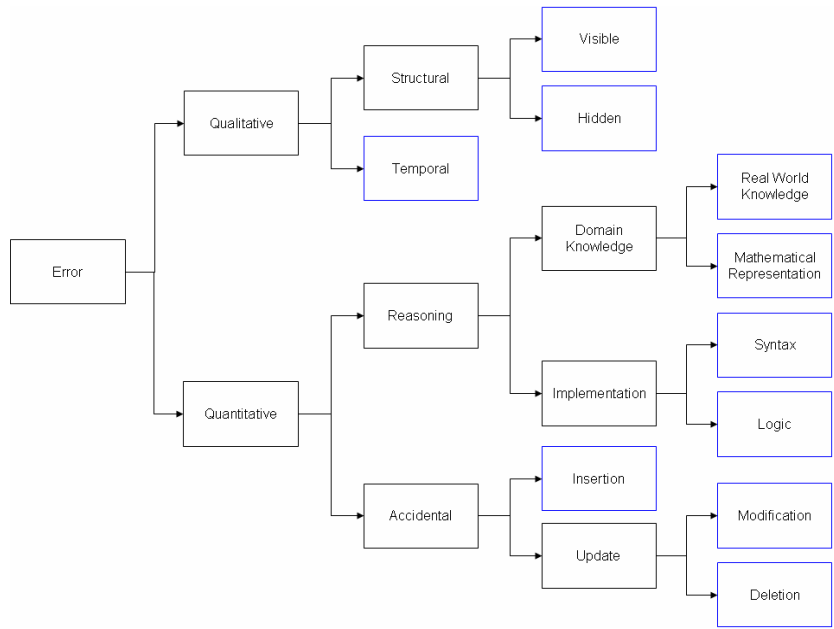

**Figure 2.2: A revision of Rajalingham's 2005 revised classification of spreadsheet error-types.**

By avoiding this distinction between end-user and developer created errors, the repetitious nature of these error-types is removed and the ambiguity of describing the error is mitigated.

The effect of removing these ambiguities shows, in figure 2.2, how ten distinct finite classifications of error-types are established, rather than the thirteen classifications containing two repetitious groups.

Over the past decade the classifications of error-types has become more comprehensive. But little has been done to investigate whether an awareness of these error-types aids spreadsheet users in identifying and consequently rectifying spreadsheet errors. This, together with the increasing availability of automated audit tools may be one reason why errors in spreadsheets still abound.

## 3    ERROR TYPE AWARENESS - THE INVESTIGATION

### 3.1    Objectives and Research Questions

The aim of the investigation was to determine if error-type awareness aids the end-user in identifying spreadsheet errors. In order to provide an effective measure of the user's ability two measures were taken. The first measure was a measure of the subject's ability to identify spreadsheet errors. The second measure was a measure of the subject's ability to identify spreadsheet errors when provide with a spreadsheet error-type taxonomy. Through comparison of these measures it was possible to determine the validity the following hypotheses:

> *[H1] An awareness of spreadsheet error-types aids the user in identifying spreadsheet errors.*
>
> *[H2] An awareness of spreadsheet error-types aids the user in rectifying spreadsheet errors.*

For the purpose of the investigation; 'identifying an error' was defined as correctly indicating the cell containing the error and naming the error-type.

In addition to the primary investigation of the above hypotheses, demographic information relating to spreadsheet experience was solicited from the subjects. Through analysis of the subjects' responses it was possible to determine the validity the following hypotheses:

> *[H3] The spreadsheet user and the spreadsheet developer is, in the majority of cases, the same person.*
>
> *[H4] The greater the experience of Microsoft Excel, the greater the success the subject will have identifying errors.*

### 3.2    Error-Types under Investigation

To investigate user awareness of error-types it was necessary to establish those error-types that were suited to the investigation. Figure 3.1 shows the chosen classification and chosen error-types at the lowest level of the taxonomy (outlined in red).

The error-types excluded from the investigation were:

- Real World Knowledge errors – as this represents the group that includes using a formula specifically for a use it has no capability of executing successfully.

- Mathematical Representation errors – as this represents the group that includes functions that require an understanding of mathematical principles such as statistical functions.

- Syntax errors – as this represents the group that includes misspelling of function names.

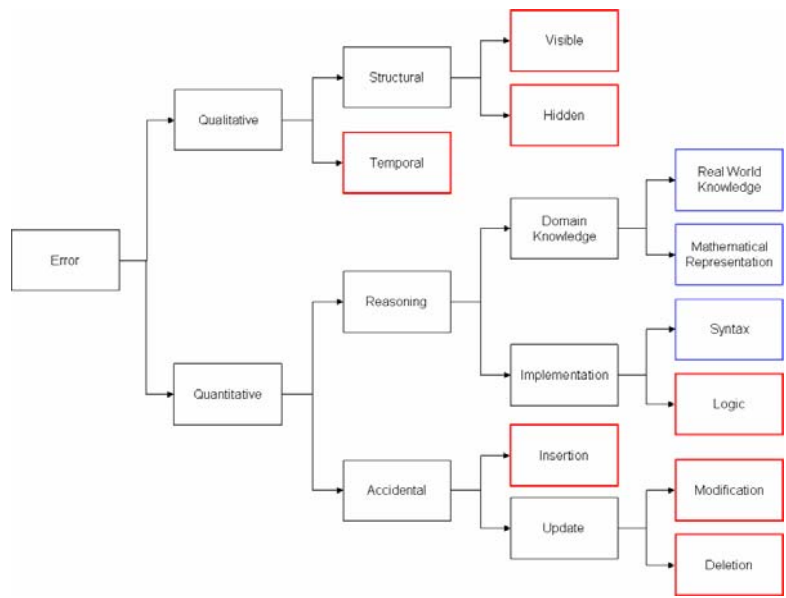

**Figure 3.1: The classification of spreadsheet error-types to be investigated.**

The first two of these error-types are based on domain knowledge, and as this is not the same for all spreadsheet users it would be erroneous to include them, as some users would have an advantage over others. The remaining error-type is automatically detected by Excel and would therefore provide little useful data.

The error-types included in the investigation were:

- Insertion errors – as this represents the group that includes omissions, duplications, and typos.

- Modification errors – as this represents the group that includes overwriting values or incorrect modifying a formula.

- Deletion errors – as this represents the group that includes erasing values or formula.

- Logic errors – as this represents the group that includes using absolute and relative references, or inserting a row of into range that is summed such that the sum does not include the new value.

- Temporal errors – as this represents the group that includes values or formulae that are accurate only for a given period.

Page 193

- Structural Hidden errors – as this represents the group that includes errors that require an examination of formulae such as hard-coded values in formulae arguments.

- Structural Visible errors – as this represents the group that includes errors that do not require an examination of formulae.

### 3.3 The Survey

The surveys comprised two web-based surveys. The first measured a baseline ability to identify errors. The second measured the ability to identify errors where error-types information was provided. In both surveys, participants were presented with a domain knowledge free and gender neutral error-seeded spreadsheet model. The participants were asked to:

- correctly identify the cell containing the error, and

- appropriately name the type of error,

for an undisclosed number of errors. In each case the model was supported with a written scenario.

### 3.4 Participants

Consistent with experimental design, a test group and a control group was surveyed. In survey 1, the test group consisted of professionals – in this sense – persons in full-time employment whose work involves the use of Excel spreadsheets. The control group consisted of students in the business and computing and mathematics schools of the University of Greenwich.

The survey was distributed to an initial selected test group of professionals and a control group of students. The test group survey participants were then asked to forward the survey to other professionals whom they believe were suitable as participants. This provided a degree controlled, random selection consistent with survey sampling protocols.

### 3.5 Procedure

The first survey was distributed via a hyperlink in an explanatory email to participants. The test and control groups emailed were large enough (greater than 20) to provide a possible quantitative number of responses.

Responses to the survey were recorded in Microsoft Access databases. The recorded responses were anonymous, unless participants requested feedback on the measurement of their error identifying ability.

The feedback also contained an explanation of the errors in the seeded model and an explanation of the error-type classification under investigation.

Requested feedback provided a mechanism for targeted distribution of the second survey. This provided the opportunity to measure both group performance and individual performance across both surveys. The control group for survey 2 was sent the survey via a hyperlink in an email to an initial group of professionals and students, excluding the feedback group of first survey. Both groups were requested to chain the email to other prospective participants in order to solicit a quantitative number of responses.

Figure 3.2 below shows the survey procedures, diagrammatically, with the number of participants indicated at each stage of the survey.

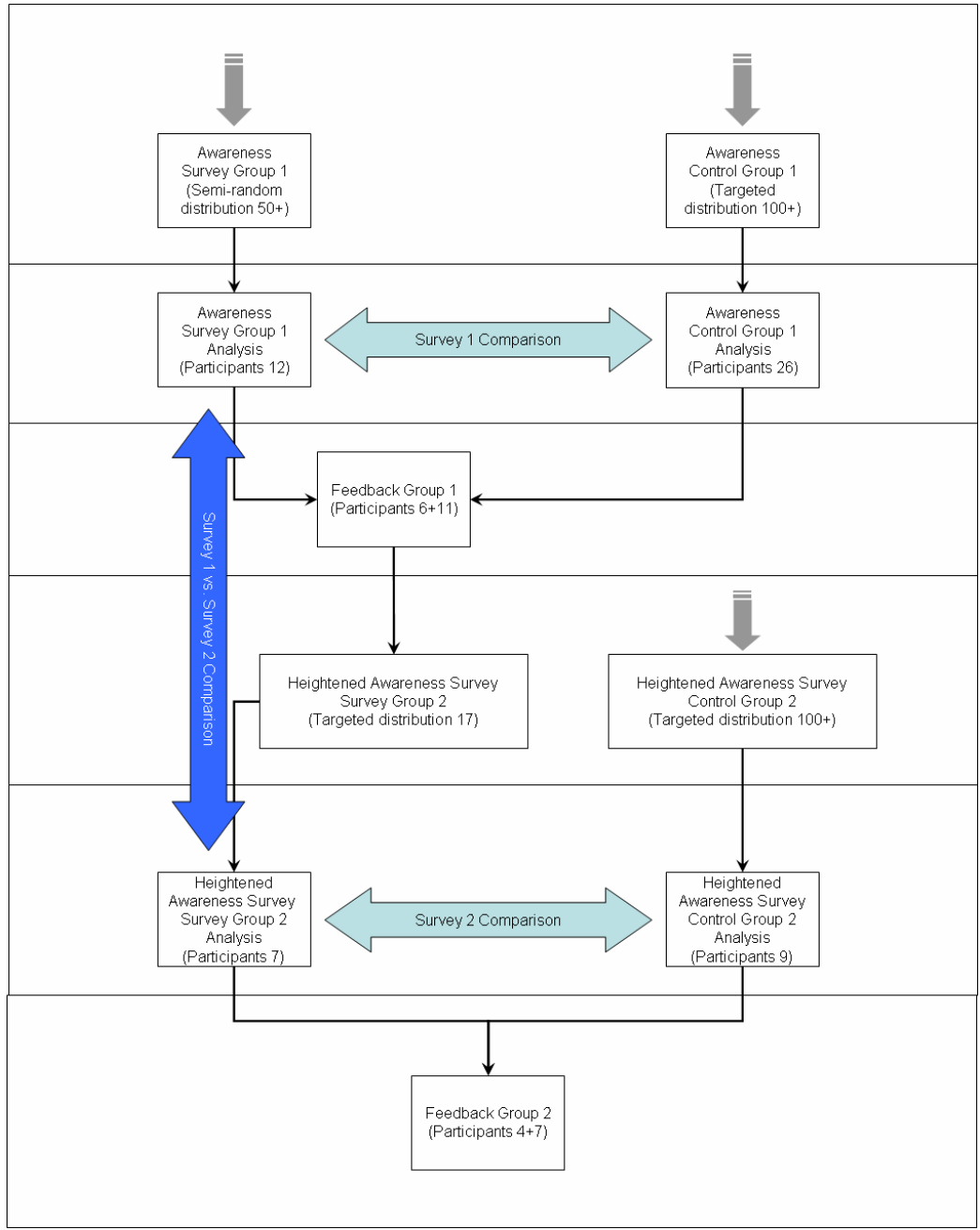

**Figure 3.2: Survey procedural flow diagram.**

### 3.6 The Survey Questionnaire and Error-Seeded Models

An established spreadsheet model was presented to the participant that had been seeded with errors. The spreadsheet model did not require any specific domain knowledge (e.g. accounting expertise) and only basic calculations such as that the area of the wall (given length and height); labour costs (given manpower and wage rate) and material costs were required. The participant was then asked to identify those errors.

### 3.7 Counting Methodology

The seeded model contained 7 errors, one for each error-type under investigation. The errors were counted in two ways, firstly by cell identification and secondly by error-type.

Errors identified by cell location multiple times regardless of error-type attributed were counted only once. Where both incorrect and correct error-types were attributed, only correct error-types were counted.

Additionally, if a participant submitted a survey response where no errors were identified it was assumed the participant failed to participate in the survey, and the results are discarded. This decision was reached after a contacting a sample of participants and asking why they unable to identify any errors. The general response was that they gave up once they realised the survey was not a simple 'pick an option' survey.

## 4  RESULTS & DISCUSSION

Consideration and discussion of the results will be conducted in the following manner:

(i) Demographic data collected from participants of survey 1.

(ii) Survey 1 test group comparison with survey 2 test group.

Table 4.1 shows the number of participants for each of the two surveys, by group (survey and control).

|               | Survey 1 | Survey 2 |
|---------------|----------|----------|
| Test Group    | 12       | 7        |
| Control Group | 26       | 9        |
| Total         | 38       | 16       |

**Table 4.1: The number of participants within each survey, by group.**

The number of participants across the two surveys (54) provided for a quantitative analysis of the responses. Again, between the surveys, the participant numbers of 38 and 16 provided for a quantitative analysis of the responses. However, it is noted that the number of participants within the groups of survey 2 were insufficient to provide a legitimate quantitative analysis.

Therefore, the following caveats were applied in consideration of the data; the analysis of demographic data for survey 1 [item (i) above], and inter-survey analysis of data for survey 1 and 2 [item (ii) above] was conducted in a quantitative mode.

### 4.1  Demographic Data Collected from Participants of Survey 1.

Demographic data was collected only during survey 1.

Table 4.2 shows the distribution (%) of participants by level of experience and the arithmetic average weekly use of Microsoft Excel. The distribution was found to be that expected between the two groups of survey 1. For example, the professionals of the test group showed that approximately 2 out 3 users had more than 6 years experience with Microsoft Excel. Whereas, the control group showed that approximately 2 out of 3 users had less than 5 years experience with Microsoft Excel. Also, the weekly average usage for test group was more than 4 times greater than that of the control group.

|  | Test Group (Professionals) | Control Group (Students) |
|---|---|---|
| Mean Weekly Usage (hrs) | 3.5 | 15.3 |
| Level of Experience |  |  |
| New (less than 2 years) | 0% | 15% |
| Short Term (2 - 3 years) | 0% | 31% |
| Medium Term (4 - 5 years) | 33% | 23% |
| Long Term (6 - 10 years) | 42% | 27% |
| Veteran (10+ years) | 25% | 4% |

**Table 4.2: Microsoft Excel average weekly usage and level of experience.**

Table4.3 shows the distribution of participants by their use of the Formula Insert and Formula Audit tools featured in Microsoft Excel. The distribution indicates that 58% of the test group and 77% of the control group use the Formula Insert tool. Clearly the majority of the control group use this tool to support the creation of formulae, while almost half of the test group appear to be confident with entering formulae directly into the spreadsheet, unsupported.

The table also indicates that both the test group and the control group use the Formula Audit tool approximately 15% less than the Formula Insert tool. As both tools may be used to audit formulae the differential in these figures may be due in part to the step-by-step nature of the Formula Insert tool. That is, the Formula Insert tool is easier to comprehend than the symbolic nature of the Formula Audit tool.

|  | Test Group | | Control Group | |
|---|---|---|---|---|
|  | Basic User | Expert User | Basic User | Expert User |
| Perceived User Type | 58% | 42% | 77% | 23% |
| Use Formula Insert Tool | 33% | 25% | 54% | 23% |
| Use Formula Audit Tool | 25% | 25% | 50% | 12% |

**Table 4.3: Survey 1 participants perceived user-type and use of Formula Insert and Audit tools.**

Table 4.4 shows two interesting pieces of data. Firstly, that 2 out of 3 participants in the control group believe that they are using error-free spreadsheets, while this figure is more than doubled in the test group (Professionals), where 5 out of 6 believe they are using error-free spreadsheets. The second interesting finding is that, in both groups, all those participants that consider themselves an expert user develop the spreadsheets they use. While, the volume of participants who consider themselves a basic user, developing the spreadsheets they use, is virtually identical for both groups – over 5 out of 6 participants.

|  | Test Group | | Control Group | |
|---|---|---|---|---|
|  | Yes | No | Yes | No |
| Use error-free spreadsheets | 83% | 17% | 65% | 35% |
|  | Basic User | Expert User | Basic User | Expert User |
| Develop own spreadsheets | 86% | 100% | 85% | 100% |

**Table 4.4: Participants perceived use of error-free spreadsheets and level of self-developed spreadsheets.**

The results from survey 1 show that error identification ability improves with experience, as can be seen in figure 4.1. Though the user categories are not linear in relation to time increments, it could be suggested that after 4 to 5 years experience with spreadsheets the error identification ability significantly increases year on year. However, to obtain a true picture of how experience improves error identification, more detailed and specific data would be required.

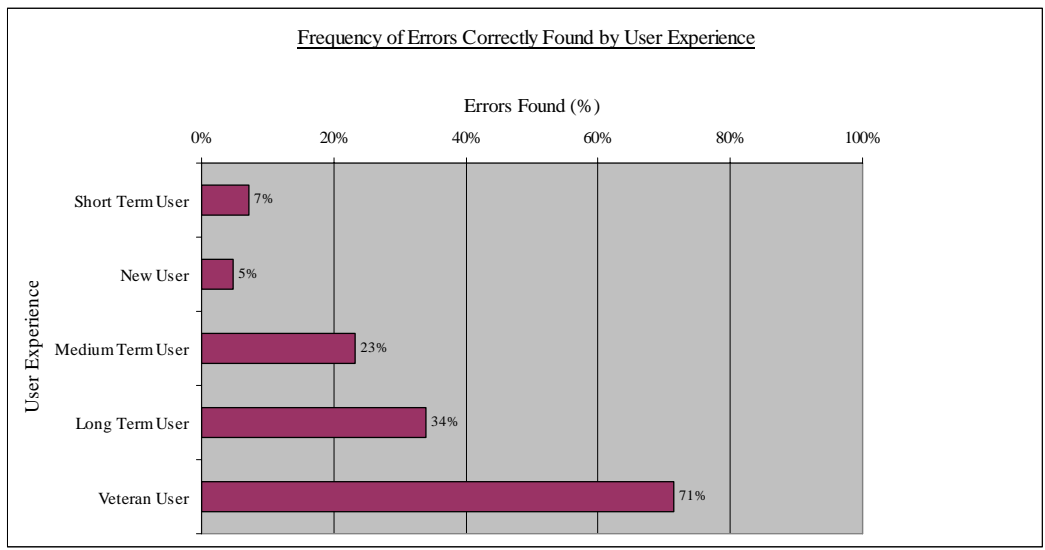

**Figure 4.1: Histogram showing the frequency errors correctly found in the error-seeded model of all participants in survey 1, by user experience.**

### 4.2 Comparison of Results of the Survey 1 Test Group with Survey 2 Test Group

The following section discusses the comparison of responses of survey 1 test group with survey 2 test group (survey 1 participants). The control groups of surveys 1 and 2 were not considered, as they had served their purpose, which was to provide a consisted differential in error identification ability across both surveys.

**Comparison of Frequency of Errors Found**

Overall the frequency of errors correctly found by the test group in survey 1 was 48%, while in survey 2 the same measure was 41%. This initially tends to suggest the classification information was of no benefit to the participants. To establish if this statement was true, it was necessary to determine if the classification had had any effect upon qualitative and quantitative error-types.

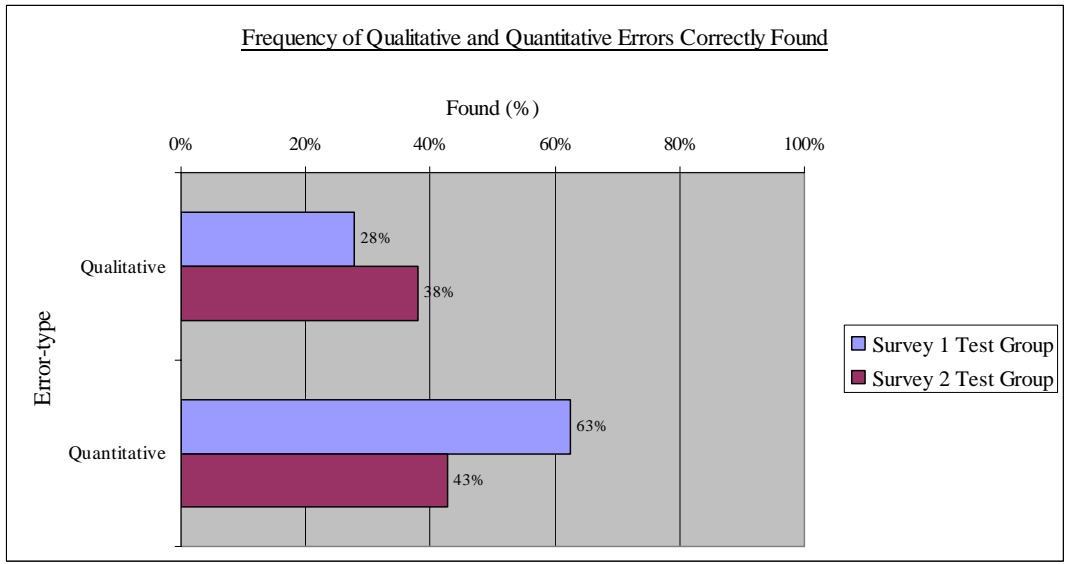

**Figure 4.2: Histogram showing the frequency of qualitative and quantitative errors correctly found by the survey groups in the error-seeded models of surveys 1 and 2.**

The histogram shown in figure 4.2 shows the classification information may have resulted in an increase in the identification frequency of qualitative errors. The value in survey 1 rose from 28% to 38% in survey 2, an increase by approximately 1/3$^{rd}$. However, there was an almost equal and opposite effect on the identification frequency of quantitative errors. The value in survey 1 fell from 63% to 43% in survey 2, a decrease by approximately 1/3$^{rd}$. This raised a question: is it possible that classification information may have a negative effect on the identification frequency of errors? This question is considered in the conclusion.

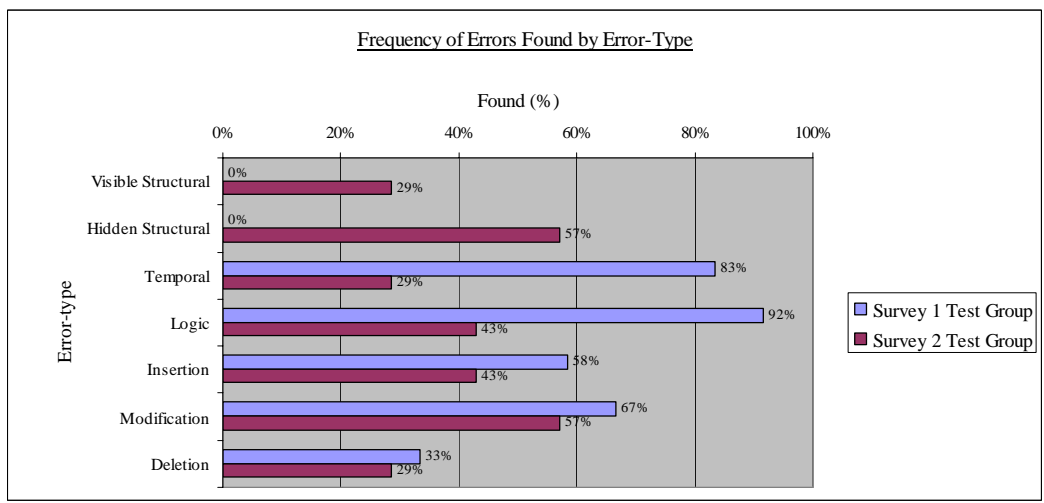

**Figure 4.3: Histogram showing the frequency of errors correctly found by the survey groups in the error-seeded models of surveys 1 and 2, by error-types.**

On a more detailed examination of the comparison by base error-types, shown in figure 4.3, the results suggested all quantitative errors had been negatively affected by the inclusion of supporting classification information. Additionally, it was discovered the temporal qualitative error identification frequency had been significantly reduced between survey 1 and survey 2.

**Comparison of distribution of Errors Correctly Found and Correctly Named**

To further understand the effects of the error-type classification information, it was necessary to determine the distribution of the response that correctly identified the errors by the survey group in both surveys 1 and 2. The box and whisker diagrams shown in figure 4.4 clearly indicate the level of correct responses deteriorated in survey 2 where supporting error-type classification information was provided.

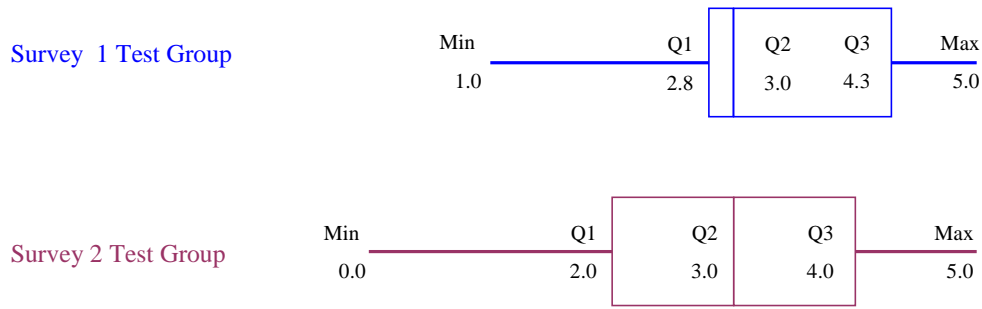

**Figure 4.4: Box and whisker diagrams showing the distribution of the actual number of errors found by the survey groups in the error-seeded models of survey 1 and 2.**

Although the error identification frequency deteriorated between survey 1 and 2, the ability to correctly name those errors identified rose, as can be seen in figure 4.5. This tends to suggest that the error-type classification information was not entirely detrimental in identifying errors. This, in turn, suggests that there may have been other influencing factors involved in reducing the error identification frequencies. Suggestions as to the possible influencing factors that may have given rise to these results are discussed in the conclusion.

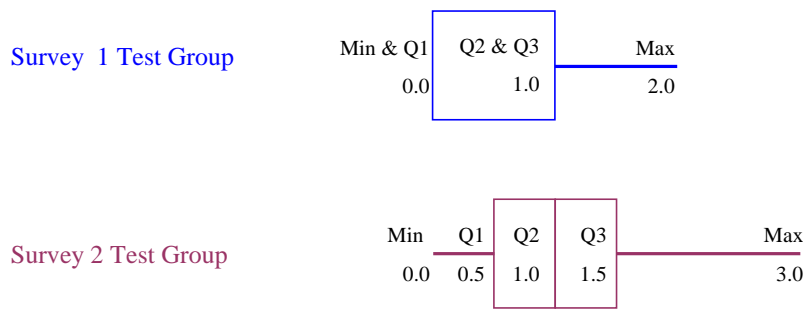

**Figure 4.5: Box and whisker diagrams showing the distribution of the actual number of errors correctly named by the survey groups in the error-seeded models of survey 1 and 2.**

### 4.3    Anecdotal Information relating to Surveys 1 and 2

In feeding back participants ability to identify errors, comments on the conduct and suitability of the surveys, and their real world representation of 'finding errors' were elicited. The following themes were consistent in participants' comments;

> *"… some errors were easy to find, but having to match the model to the scenario made finding the others more difficult …"*

> *"… you know instinctively if it is right or wrong, but I wouldn't have previously thought about how to classify them …"*

> *"... although I use Excel on a regular basis ... I had no idea about the "types" of errors ...*
>
> *"... checking for errors is easy if you're used to auditing ..."*
>
> *"... I find the best way to help ensure [spreadsheets] are correct is to have built in sub total checks that, for example compare results calculated using different methods ..."*
>
> *"...[I create] models that [have] some reconciliation and checking for inconsistencies ..."*
>
> *"... finding errors in spreadsheet depends on how complex the spreadsheet is ..."*
>
> *"... a lot of the spreadsheets we work on deal with linked files, large volumes of data often from different sources, and complex calculations ..."*

These themes of: understanding the context of a model, not considering error classification, audit technique familiarity, reconciliation and inconsistency checking, model and formula network complexity, and data source awareness suggest there are other factors heavily influencing the ability of the user to identify errors in spreadsheets.

These comments also suggest that spreadsheet users are aware that some form of error checking is required in order to maintain the integrity of spreadsheets. However, the distinct impression was given that, which method is applied, is dependent on the user.

## 5     CONCLUSIONS

The research presented in this dissertation clearly shows there has been limited research to establish a classification of spreadsheet errors. The research that has been carried out has produced viable classifications, furthering development of a common classification and providing valuable support for future research.

The investigation set out to establish the validity of a number of hypotheses:

> *[H1] An awareness of spreadsheet error-types aids the user in identifying spreadsheet errors.*

Although the results show a minor degradation in the overall error identification frequency, it cannot be conclusively determined whether the hypothesis was proved or disproved. However, as the results did clearly show qualitative error identification had improved, it can be concluded that awareness of spreadsheet error-types aids the user in identifying qualitative spreadsheet errors. And thus, logically awareness of spreadsheet error-types should aid the user in identifying quantitative spreadsheet errors.

> *[H2] An awareness of spreadsheet error-types aids the user in rectifying spreadsheet errors.*

Although the results do not allow determination of proof or disproof, it is logical to suggest that increased identification provides the potential for increased error rectification, as without first identifying errors they may not be rectified.

> *[H3] The spreadsheet user and the spreadsheet developer is, in the majority of cases, the same person.*

The results clearly prove the hypothesis. Over 85% of the survey participants claimed development of the spreadsheets they use. This provides for serious consideration of the revised classification (see figure 3.1) presented in this research as a simplified classification that removes the need to consider the 'when and who' of error creation.

> *[H4] The greater the experience of Microsoft Excel, the greater the success the subject will have identifying errors.*

The results clearly prove the hypothesis. The results presented in this research (see figure 4.1) clearly showed a correlation between error identification ability and user experience. However, further consideration should be given to this hypothesis, as the user experience did not consider average weekly time spent using Microsoft Excel. This is an important factor that should not be overlooked in any similar research in the future.

Overall no firm conclusion one way or the other may be drawn from the results to provide an answer to the research question:

> *"Does an awareness of differing types of spreadsheet errors aid end-users in identifying spreadsheets errors?"*

However, the error classification information provided with survey 2's model did aid the participants in identifying more qualitative spreadsheet errors than in survey 1. The same is not true of the identification of quantitative spreadsheet errors.

The results of the surveys and anecdotal information clearly shows there are many negative factors which make error identification difficult. Factors such as: model and formula network complexity, a strong belief that the spreadsheet in use is error-free, a lack of commitment to audit techniques, and spreadsheet development by self- and un-trained users.

However, there are also a number of positive factors which aid error identification. User experience and training, a commitment to audit technique, use of an audit tool, knowledge of the model's context, data source awareness, and the model design, are just some of positively influencing factors.

This investigation has shown that error classification awareness overall must be added to the list of positive factors that aid in spreadsheet error identification.

### 5.1    Limitations

The volume of participants in the investigation limited the ability of the researcher to quantifiably conclude that error classification awareness aids spreadsheet error identification. The simplicity/complexity of the survey models and the unsupervised survey participation limited the control of other influencing factors such as volume of participants and commitment to the research by participants.

### 6    FURTHER RESEARCH

The research carried out in this investigation has looked at a small specific area of the problem of errors in spreadsheets. In order to fully understand the problem of errors this work requires extension. This investigation should be re-scaled in order to obtain more data to definitively prove or disprove the overriding question of whether error-type awareness aids in spreadsheet error identification.

Having determined if error-type awareness is of benefit it would be useful to establish what type of organisations are prolific spreadsheet users, and who within the organisation uses spreadsheets and for what purpose.

These two pieces of research would then provide an excellent basis for the examination of how the level of user training and experience impacts on the level of error identification. The aim would be to determine if positive behavioural change affects the user's ability to identify and avoid errors.

Increasing reliance on automatic error identification and spreadsheet audit suggests that a full examination of the approaches employed by these tools should be made. This work should also involve a consideration of how effective they are at identifying errors when compared to the average user employing appropriate audit techniques.

Underlying the need for this research is the continuous drive for more sophisticated and complex spreadsheet models. An examination of best practice approaches would help to establish the determining factors in spreadsheet model design that aid in avoiding error creation.